\documentclass[11pt]{article}
\usepackage[dvips]{graphicx}  
\begin{document}

\title{\LARGE \bf Process Physics:
 Modelling Reality as Self-Organising Information\footnote{Preview of {\it  Process Physics:
The Limits of Logic and the Modelling of Reality}; in preparation.}}  
\author{{Reginald T. Cahill, Christopher M. Klinger and Kirsty Kitto}\\
  {School of Chemistry, Physics and Earth Sciences}\\{ Flinders University
}\\ { GPO Box
2100, Adelaide 5001, Australia }}

\date{}
\maketitle

\begin{abstract}
 The new {\it Process Physics} models reality as self-organising relational information and takes
account of  the limitations of logic,  discovered by G\"{o}del and extended by Chaitin, by using the concept of
self-referential noise.   Space and quantum physics are emergent and unified, and described    by a Quantum 
Homotopic Field Theory of fractal topological  defects embedded in a three-dimensional fractal process-space. 
  
\end{abstract}

\newpage

\noindent {\large \bf  Modelling Reality}  

The present day modelling of reality and the mindset of physicists was very much set by the Ancient
Greeks some 2,500 years ago, particularly by Democritus with his concept of atoms as  objects occupying a
position in space.  Ever since  physicists  have believed in objects and their {\it a priori} rules of behaviour
or `laws of physics' as fundamental to modelling reality. This mode of modelling   has been extremely
successful. These concepts were clearly abstractions from everyday human experience and culminated, in the case
of space, with the Euclidean formalisation of  geometry. Great progress followed
 Galileo's and then Newton's demonstrated  successes in using a geometrical model of the phenomena of
time, despite the glaring deficiencies of that model,  which matches the
ordering of events with the ordering of the real numbers,  but fails to find a match for the contingent
present moment or even the difference between past and present.  These were serious problems that
persisted in the more elaborate spacetime geometrical model  by Einstein and were of great concern
to him, but even Einstein doubted that any modelling of time could capture the present moment effect.
Significantly, because this static real-number modelling of time fails in these respects, one must
always introduce a meta-rule which states that, in essence, one must imagine a point moving along the
time-line at a uniform rate; something we always do subconsciously as physicists. We should  call
this the geometrical-time meta-rule.  It is important to note that in Newtonian physics and
all that followed the modelling of reality is actually {\it non process}:  there is no sense of
anything actually happening. Newton's celebrated equations of motion don't actually describe motion;
they describe {\it static} functions,
$x(t)$ say, which we relate to motion by using the geometrical-time meta-rule.

An even greater problem for reality modelling arose with the quantum theory. There static
or non-process wave-functions  $\psi(x,t)$  were related to motion of the wavefunction by invoking the
geometrical-time meta-rule, but that was manifestly inadequate since the clicking of detectors
clearly revealed an additional  real process that was completely absent  in the quantum theory.  This 
 mismatch was `fixed' by Born  introducing the  quantum  measurement meta-rule that states that the
probability density of an actual click at location $x$ in a detector at time $t$ is given by $|\psi(x,t)|^2$.
Physicists appear to believe   that these clicks are to be understood as being produced by
objects (`particles'), that are accompanied by  $\psi(x,t)$, hitting the detector  even though these objects are
not mentioned in the quantum mathematical formalism, which deals only with $\psi(x,t)$, say.  

The success of physics thus actually arises, in part,  from its meta-rules. The need
for them, being separate from  but consistent with  the  mathematical formalisms,   are 
indicators of a deep flaw within the current mindset of physicists.  An analogous deep flaw in the
axiomatic formalisms  of mathematics was revealed by G\"{o}del in 1931.  He showed that for  formalisms
(arithmetic initially)  sufficiently rich that they support self-referential statements,   there
exist truths which cannot be proven within the formalism. Chaitin\footnote{G.J. Chaitin, {\it The
Unknowable}, Springer-Verlag, Germany 1999, \newline
http://www.cs.auckland.ac.nz/CDMTCS/chaitin/unknowable/index.html}, more recently showed that these unprovable
truths have the property of randomness; from the point of view of the given formalism they have  no
explanation.  In physics we would describe them as un-caused.   While physics has never reached the stage of a
strict axiomatic formalism, it nevertheless has been travelling in that direction and, like mathematics, there
are truths beyond formalism. The absence of such contingent truths  has been partialy compensated by the use of
meta-rules.   

In process physics  we  present a  radical new modelling of reality designed to overcome these and other
deep problems within current physics. The key concept is to  model reality as  self-organising relational
information  via an order-disorder  process system which takes account of the limitations of
formalism or logic by using the new concept of self-referential noise (SRN).  This SRN is not randomness due
to a simple lack of potential knowledge (such as found in statistical physics) but mimics what are {\it in
principle} un-provable or non-algorithmic  truths; by using SRN we essentially transcend the limits of
logic without being illogical. Logic, it should be noted, is the language of named `objects'; for this reason we
start up process physics using `pseudo-objects'. The dramatic discovery is that rather than being some 
impediment to understanding reality G\"{o}del's discovery and its extension to SRN acts as an intrinisic
resource within this non-formal system and with which  a vastly improved modelling   of reality  has become
possible.   By  inducing, in approximation,  a formal system together with associated emergent meta-rules it
links back to and subsumes the current physics modelling of reality.  We find that the system operates by
forming a dissipative structure, driven by the SRN, and  which is characterised by an emergent and expanding
three-dimensional fractal process-space in which are embedded  self-replicating fractal topological defects,
both described in a unified manner by a Quantum Homotopic Field Theory (QHFT). This emergence is a
non-algorithmic increase in complexity in the system. 
  
The process modelling of reality dates back to Heraclitus of Ephesus (540-480BC) who argued that common sense is
mistaken in thinking that the world consists of stable `things'; rather the world is in a state of flux and the
appearance of `things' depend upon this flux for their continuity and identity. So process physics  has also
been described as a Heraclitean Process System, with the flux identified with SRN.

\vspace{3mm}
\noindent {\large \bf Relational Information and Self-Referential Noise}  

Process physics   models reality as a self-organising relational information 
system,  sufficiently rich that self-referencing is possible (`relational information' refers to 
the idea that information is internal to the system). Curiously  a similar task arises in
modelling  consciousness.    For such a system G\"{o}del's key discovery was that truth has no finite
description, and we model this, borrowing from Chaitin's work in mathematics, by introducing the concept of an
intrinsic randomness which  is called self-referential noise (SRN).  SRN ensures that most truths are contingent
- they are un-caused;  a restricted form of  determinism is then an emergent feature of process physics.   

Because process physics is at its deepest
levels an information system and is devoid of objects and their laws it requires a subtle bootstrap mechanism to
set it up:  we introduce  real-number valued connections or relational information strengths $B_{ij}$  between
any two pseudo-objects $i$ and $j$ (also called  monads after Leibniz who espoused the relational mode
of thinking in response to and in contrast to Newton's absolute, ie objective, space and time). These
pseudo-objects are to be regarded as temporary scaffolding. They will be revealed  to be themselves sub-networks
of informational relations. To avoid explicit self-connections $B_{ii}\neq 0$, which are a part of the
sub-network content of
$i$, we use antisymmetry $B_{ij}=-B_{ji}$ to conveniently ensure that 
$B_{ii}=0$. At this stage a key concept of process physics arises:  to ensure that the monads are not objects
the system must generate  linked  monads forming a fractal network; then
self-consistently the start-up monads may themselves be considered   as mere names for sub-networks of relations
(for a successful suppression  the scheme must display self-organised criticality). 

To
generate a fractal structure we must  use a non-linear iterative system for the
$B_{ij}$ values.  These iterations amount to the logical necessity to introduce a time-like phenomena into
process physics.   Any system possessing {\it a priori}  `objects' can never be fundamental as the
explanation of such objects must be outside  the system.  Hence in process physics the absence of intrinsic
undefined objects is linked with the phenomena of time. In this way process physics arrives at a new modelling of
time, {\it process time}, which is much more complex than that introduced by Galileo and reaching its high point
with Einstein's spacetime geometrical model. In process physics Einstein's model emerges not as some ontological
statement about reality, such as ``the universe is 4D geometry'', but as the coarse-grained `history book' of the
evolving $B_{ij}$ network. Geometrical time is then merely the pagination of the history book, and this has no
ontological reality. The nature of the internal experiential time is complex but  we expect time dilation and
other manifestations  through  the  Lorentzian  explanation; that is, the system mimics covariance effects but
is not intrinsically covariant. Such effects are caused by the finite information processing rate.

The process physics  concepts have so far only been realised with one particular  scheme involving a
non-linear matrix iteration with additive SRN $w_{ij}$;
\begin{equation}
B_{ij} \rightarrow B_{ij} -\alpha (B^{-1})_{ij} + w_{ij},  \mbox{\ \ } i,j=1,2,...,2M;
M
\rightarrow
\infty.
\label{eq:map}\end{equation}
 The
 $w_{ij}=-w_{ji}$ are
independent random variables for each $ij$ pair and for each iteration, chosen from some probability
distribution.  We start the iterator at  $B\approx 0$ - representing the absence of information.
  With the noise absent the iterator would converge to
 a constant matrix.
However in the presence of the noise the dominant mode is the formation of a randomised and structureless 
background. However  a significant discovery was that the noisy iterator also manifests  a   self-organising
process which results in a growing  three-dimensional  fractal process-space that competes with this random
background - the formation of a `bootstrapped universe'.

\vspace{3mm}
\noindent {\large \bf 3D Process Space  from Self-Assembling Gebits}  

 This  growing  three-dimensional  fractal process-space is an example of a Prigogine far-from-equilibrium
dissipative structure driven by the SRN. 
 From each iteration the noise term
will additively introduce rare large value
$B_{ij}$.  These  $B_{ij}$, which define sets of linked monads, will persist   through more
iterations than smaller valued $B_{ij}$ and, as well,  they become further linked  by the iterator
to form a three-dimensional process-space with embedded topological defects.
 
 To see this consider a monad $i$ involved in one such large $B_{ij}$;   
 it will be   connected via  other large $B_{ik}$ to a
number of other monads and so on, and this whole set of connected monads forms a unit which we call a gebit as
it acts as a small piece of geometry formed from random information links and  from which the
process-space is self-assembled. The gebits compete for new links  and undergo mutations. 

To analyse the connectivity of such 
gebits assume for simplicity that the large $B_{ij}$  arise with fixed but very small probability $p$, then the
geometry of  the gebits is revealed by studying the probability distribution for  the structure of the
gebits minimal spanning trees with $D_k$ monads  at $k$ links from monad $i$ ($D_0 \equiv 1$), this is given
by\footnote{G. Nagels, {\it Gen. Rel. and Grav.} {\bf 17}, 545 (1985).
}
$${\cal P}[D,L,N] \propto \frac{p^{D_1}}{D_1!D_2!....D_L!}\prod_{i=1}^{L-1}
(q^{\sum_{j=0}^{i-1}{D_j}})^{D_{i+1}}(1-q^{D_i})^{D_{i+1}},$$
where $q=1-p$, $N$ is the total number of monads in the gebit and $L$ is the maximum depth from monad $i$. 
To find the most likely connection pattern we numerically maximise ${\cal P}[D,L,N]$ for fixed $N$ with respect
to
$L$ and the $D_k$. The resulting $L$ and $\{D_1,D_2,...,D_L\}$ fit very closely to the form $D_k\propto
\sin^{d-1}(\pi k/L)$;  see fig.1a  for $N=5000$ and $\mbox{Log}_{10}p=-6$.  The resultant  $d$ values for
a range of $\mbox{Log}_{10}p$ and $N=5000$ are shown in fig.1b. 

\begin{minipage}[b]{75mm} 
\includegraphics[ scale=0.89]{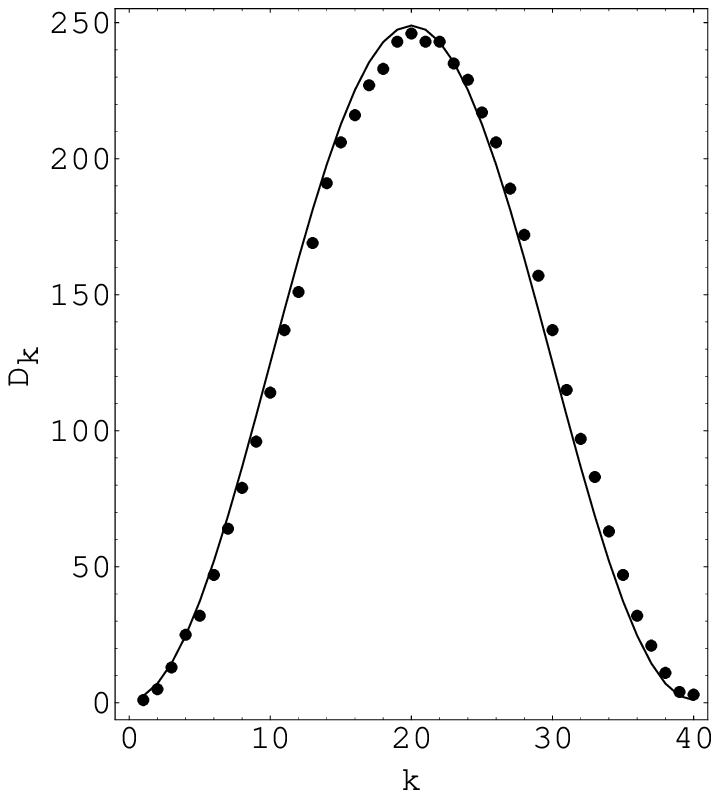}
\makebox[75mm][c]{(a)}
\end{minipage}
\begin{minipage}[b]{50mm} 
\hspace{0mm}\includegraphics[ scale=0.97]{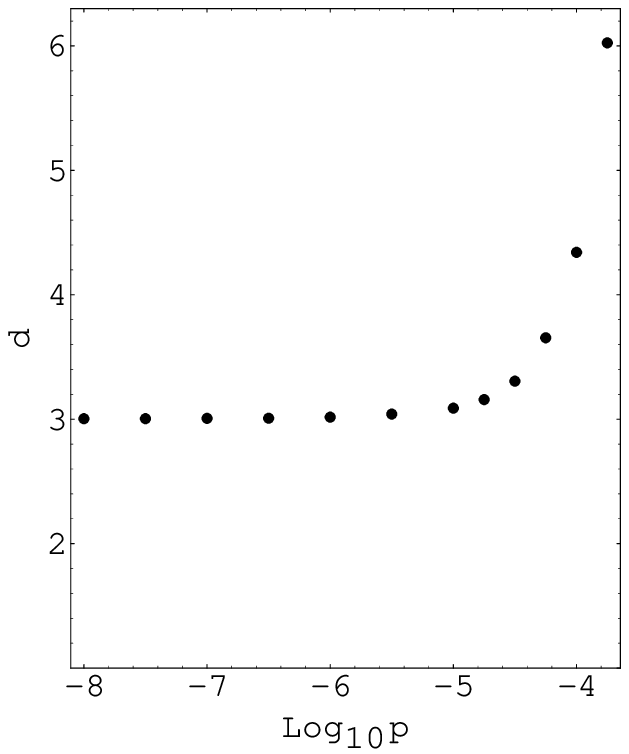}
\makebox[65mm][c]{(b)}
\end{minipage}
 
\begin{figure}[ht] 
\caption{\small 
(a) Points show the $D_k$ set and $L=40$ value found by numerically  maximising ${\cal P}[D,L,N]$
for $\mbox{Log}_{10}p=-6$ for fixed  $N=5000$. Curve shows
$D_k\propto \sin^{d-1}(\frac{\pi k}{L})$ with best fit $d=3.16$ and $L=40$, showing  excellent
agreement, and indicating  embeddability in an $S^3$ with some topological defects. (b) Dimensionality $d$ of
the gebits as a function of  the probability $p$.} 
\end{figure}

This shows, for $p$ below a critical value, that
$d=3$ indicating that the connected monads have a natural embedding in a 3D hypersphere $S^3$; call this
a base gebit. Above that value of $p$,   the increasing value of $d$
indicates the presence of extra links that, while some conform with the embeddability,  are in the main defects
with respect to the geometry of the $S^3$.  These extra links act as topological defects.  By themselves these
extra links will have the   connectivity and embedding geometry of numbers of gebits, but these gebits have a
`fuzzy' embedding in the base gebit. This is an indication of  fuzzy  homotopies  (a homotopy is,
put simply, an embedding of one space into another).  

The base gebits $g_1, g_2, ...$ arising from the SRN together with their embedded topological defects have
another remarkable property:  they are `sticky' with respect to the iterator.  Consider the   larger valued 
$B_{ij}$ within a given gebit  $g$, they form  tree graphs and most tree-graph adjacency matrices are singular (
det$(tree)=0$).  However  the presence of other smaller valued $B_{ij}$ and the general
 background noise  ensures that det$(g)$ is small  but not exactly zero. 
Then the 
$B$ matrix has an inverse with large components that act to cross-link  the new and
existing gebits.  If this cross-linking was entirely random then the above analysis  could again be used and we
would conclude that  the base gebits themselves are formed into a 3D hypersphere with embedded topological
defects.  The nature of the resulting 3D process-space is suggestively indicated in fig.2.  

\begin{figure}[ht] 
\hspace{26mm}\includegraphics[scale=0.8,angle=90]{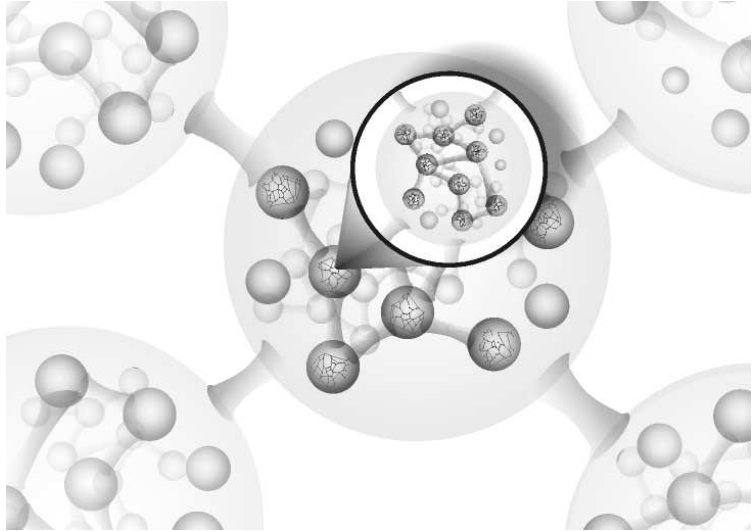}
\caption{\small  Artist representation (including magnifying glass) of linked and embedded gebits forming
a 3D fractal process-space - characterised by a quantum-foam behaviour.
 \label{figure:gebits}}
\end{figure}

Over ongoing
iterations the existing gebits become cross-linked and eventually lose their ability to undergo further
linking; they lose their `stickiness'  and decay.  Hence the emergent space is 3D but is continually
undergoing  replacement of its component gebits;  it is an informational process-space, in sharp distinction to
the non-process continuum geometrical spaces that have played a dominant role in modelling physical space.  If
the noise is `turned off' then this emergent dissipative space will decay and cease to exist.  We thus see that
the nature of space is deeply related to the logic of the limitations of logic;  by making {\it a priori}
geometrical assumptions about space physicists have inadvertently treated space as an object rather than as a
process, and a process requiring more than the geometrical modelling of time.  Next we turn to  emergent
quantum behaviour.          

\vspace{3mm}
\noindent {\large \bf Quantum Homotopic Field Theory} 

Relative to the iterator the dominant  resource is the large valued $B_{ij}$ from the SRN
because they form the `sticky' gebits which are self-assembled into the non-flat compact 3D process-space. The 
accompanying topological defects within these gebits and also the topological defects within the process space
require a more subtle description.  The key behavioural mode for those defects which are  sufficiently large
(with respect to the number of component gebits) is that their existence, as identified by their topological
properties, will survive the ongoing process of mutation, decay and regeneration; they are topologically
self-replicating.   Consider the analogy of a closed loop of string containing a knot - if, as the string ages,
we replace small sections of the string by new pieces then eventually all of the string will be replaced;
however the relational information represented by the knot will remain unaffected as  only the topology of the
knot is preserved.
  In the process-space there will be gebits embedded in gebits, and so forth,  in topologically
non-trivial ways;  the topology of these embeddings is all that will be self-replicated in the processing of
the dissipative structure.   

To analyse and model the life of these topological defects we need to characterise
their general behaviour: if sufficiently large (i) they will self-replicate if topologically non-trivial,
(ii)  we may apply continuum homotopy theory to tell us which embeddings are topologically
non-trivial,  (iii)  defects will only dissipate if embeddings of `opposite winding number' (these classify the
topology of the embedding)  engage one another, (iv)  the embeddings will be in general
fractal, and (iv) the embeddings need not be `classical', ie the embeddings will  be fuzzy.  To track the 
coarse-grained behaviour of such a system has lead us to the development of  a new form of quantum field
theory:  Quantum Homotopic Field Theory (QHFT)\footnote{There may be a connection to V. Turaev, {\it Homotopy
field theory in dimension 3 and crossed group-categories}, math.GT/0005291.}.  This models both the
process-space and the topological defects.

 QHFT has the form of a functional Schr\"{o}dinger equation for the time-evolution
of a wave-functional 
$\Psi(....,\pi_{\alpha\beta},....,t)$
$$i\hbar\Delta\Psi(....,\pi_{\alpha\beta},....,t)=H\Psi(....,\pi_{\alpha\beta},....,t)\Delta t +  \mbox{QSD
terms},
$$      where the configuration space is that of all possible 
   homotopic mappings; $\pi_{\alpha\beta}$ maps from $S_\beta$
to  $S_\alpha$ with $S_\gamma \in  \{S_1,S_2,S_3,...\}$ the set of all possible gebits (the topological defects
need not be $S^3$'s). Depending on the `peaks' of 
$\Psi$ and the connectivity of the resultant dominant mappings such mappings are to be interpreted as either
embeddings  or links; fig.2 then suggests the dominant process-space form within $\Psi$ showing both links and
embeddings. Space then has the characteristics of Wheeler's quantum foam.  We have indicated Planck's constant
$\hbar$ to emphasise that  100 years after its discovery  we finally have an explanation for its logical
necessity in describing reality.  Its actual value depends on an arbitrary choice of units;  here the natural
value is
$\hbar=1$. 

There are  additional Quantum State Diffusion (QSD)\footnote{I.C. Percival, {\it Quantum State Diffusion},  
Cambridge Univ. Press, 1998.} terms which are non-linear and stochastic; these QSD terms are ultimately
responsible for the emergence of classicality via an  objectification process, but in particular  they produce
wave-function(al) collapses during quantum measurements; a mechanism that eluded quantum theory since its
discovery and which is finally seen to have its explanation with G\"{o}del's incompleteness theorem and its
associated SRN within a process-system. The random click of the detector is then a manifestation of Geodel's
profound insight that truth has no finite description in self-referential systems;  the click is simply a
random contingent truth. The SRN is thus seen to be  a major missing component of the modelling of reality.  
In the above we have  a deterministic and unitary evolution, tracking and preserving topologically encoded
information, together with the stochastic QSD terms, whose form protects that information during
localisation events, and which also  ensures the full matching in QHFT of process-time to real time: an ordering
of events, an intrinsic direction or `arrow' of time and a modelling of the contingent present moment effect.

The mappings  $\pi_{\alpha\beta}$ are related to group manifold parameter spaces with the group determined by
the dynamical stability of the mappings, this gauge symmetry leads to  the flavour symmetry of the standard
model. Quantum homotopic mappings  behave as fermionic or bosonic modes for  appropriate winding numbers; so
process physics predicts both fermionic and bosonic quantum modes, but with these  associated with
topologically encoded information and not with  objects or `particles'.  Unlike conventional quantum field
theory QHFT has fractal embedded fermionic/bosonic modes. 

The solution of the Schr\"{o}dinger functional equation, without the QSD terms, 
can be expressed as  functional integrals; and using functional integral calculus techniques these can be
deconstructed down to  preon fermionic functional integrals but only by introducing a meta-colour dynamics,
with the colour  necessarily `confined' as there are no  topological defects corresponding to the preons.

\vspace{20mm}
\setlength{\unitlength}{2.0mm}
\begin{picture}(0,0)
\thicklines
\qbezier(5,5)(15,+4)(20,0)\qbezier(20,0)(28,-5)(35,-2)
\qbezier(5,5)(12,10)(20,10)
\qbezier(35,-2)(42,+3)(50,3)
\qbezier(27.8,9)(30,+9)(35,5)  
\qbezier(35,5)(38,2)(50,3)

\qbezier(5,4)(15,+3)(20,-1)
\qbezier(20,-1)(28,-6)(35,-3)
\qbezier(35,-3)(42,+2)(50,2)

\put(5,4){\line(0,1){1}}\put(35,-3){\line(0,1){1}}
\put(50,2){\line(0,1){1}}

\put(18,7){\circle*{2}}\put(18.5,6.8){\circle*{2}}
\put(31,1){\circle*{5.0}}\put(31.6,1.26){\circle*{5.}}


\put(18.5,8.0){\circle{2}}
\put(19.2,8.8){\circle{2}}
\put(20,9){\circle{2}}
\put(21,10){\circle{2}}
\put(21,9){\circle{2}}
\put(22.5,10){\circle{2}}
\put(23.5,9.5){\circle{2}}
\put(23.9,10.5){\circle{2}}
\put(25.2,10.6){\circle{2}}
\put(25.4,9.3){\circle{2}}
\put(26.0,9.8){\circle{2}}
\put(27.0,9.0){\circle{2}}
\put(27.5,8.0){\circle{2}}
\put(28.7,7.5){\circle{2}}
\put(29.5,6.0){\circle{2}}
\put(28.2,6.0){\circle{2}}
\put(28.5,5.0){\circle{2}}
\put(29.3,4.0){\circle{2.5}}
\put(30.3,4.5){\circle{2.5}}
\put(31.0,2.7){\circle{2.5}}
\put(30.2,2.2){\circle{2.5}}
\end{picture}

\begin{figure}[ht]
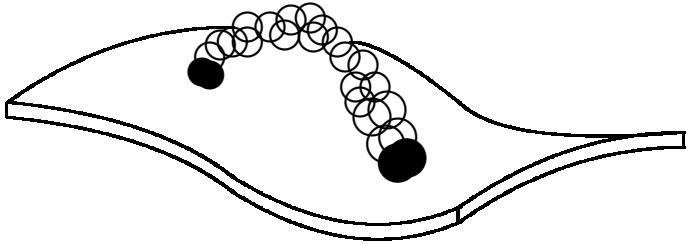
 
\vspace{5mm}\caption{\small  Topologically encoded  information non-locally embedded  in
the process-space with two `footprints' (process-space drawn as 2D with thickness and not showing gebit
structure).
 \label{figure:Footprint}}
\end{figure}

The topologically encoded information  may have more than one `foot-print' in the process-space, as indicated in
fig.3. In
  the induced approximate formal standard quantum theory they
 correspond to the superpositions  $\psi_1(x)+\psi_2(x)$. So we also finally understand
quantum non-locality as illustrated most strikingly with the two-slit experiment for photons, but also by 
EPR entanglement.   The localisation of such  states is caused by the QSD terms acting non-locally via
the macroscopic detectors, which are themselves permanently localised by QSD effects.   Process physics is a
stochastic non-local hidden variable theory, and so is consistent  with the  experimentally observed  violation
of the  Bell's inequalities for local hidden variable theories. 

Process physics is seen to realise Wheeler's suggested informational `{\sl it from bit}' program via the
sequence `{\sl bit} $\rightarrow$  {\sl gebit} $\rightarrow$ {\sl qubit} $\rightarrow$ {\sl it}', but only
by modelling  G\"{o}delian  limitations on informational  completeness at the {\sl bit} level.  Process
Physics is at the same time  deeply bio-logical - reality is revealed as a self-organising, evolving and
competitive information processing  system; at all levels reality  has evolved processes for self-replicating
information.

\vspace{3mm}
\noindent {\large \bf Process Physics Resources}            

\begin{enumerate}

\item  R.T. Cahill and C.M. Klinger, {\it Bootstrap Universe from Self-Referential
Noise}, gr-qc/9708013.

\item R.T. Cahill and C.M. Klinger, {\it Self-Referential Noise
and the Synthesis of Three-Dimensional Space}, {\it Gen. Rel. and  Grav.} {\bf 32}(3), 529
(2000); gr-qc/9812083.

\item R.T. Cahill and C.M. Klinger, {\it Self-Referential Noise
as a Fundamental Aspect of Reality},
 Proc.~2nd Int.~Conf.~on Unsolved Problems of Noise and Fluctuations
(UPoN'99), eds.~D.~Abbott and L.~Kish, Adelaide, Australia, 11-15th July
1999, {\bf Vol.~511,} p.~43 (American Institute of Physics, New York, 2000); gr-qc/9905082.

\item M. Chown, {\it Random Reality}, New Scientist, Feb 26 2000, Vol 165 No 2227, 24-28; 

http://www.newscientist.com/features/features.jsp?id=ns22273

\item C. Gent, {\it Is Reality a Side-Effect of Randomness?},  Flinders Journal,  Vol 11 No 4, 2000; 
http://adminwww.flinders.edu.au/PRIO/Journal/FJournal2000No4.html

\item R.T. Cahill, C.M. Klinger and K. Kitto, {\it  Process Physics:
The Limits of Logic and the Modelling of Reality}; in preparation.

\item Process Physics Web Page: 

http://ph131.ph.flinders.edu.au/html/people/processphysics.html 

\item The  precursor to process physics was

\noindent R.T. Cahill and C.M. Klinger, {\it Pregeometric Modelling of the
Spacetime Phenomenology}, Phys. Lett. A {\bf 223}(313)(1996); p217, Proc. First Australasian Conference on
General Relativity and Gravitation, ed. D.L. Wiltshire, Inst. Theor. Physics, University of Adelaide 1996;
 gr-qc/9605018.

\end{enumerate}

\end{document}